\documentclass[sigconf]{acmart} %

\AtBeginDocument{%
  }

\copyrightyear{2026}
\acmYear{2026}
\setcopyright{cc}
\setcctype{by}
\acmConference[DAC '26]{63rd ACM/IEEE Design Automation Conference}{July 26--29, 2026}{Long Beach, CA, USA}
\acmBooktitle{63rd ACM/IEEE Design Automation Conference (DAC '26), July 26--29, 2026, Long Beach, CA, USA}
\acmDOI{10.1145/3770743.3804125}
\acmISBN{979-8-4007-2254-7/2026/07}

\usepackage{arydshln}
\usepackage{multirow}
\usepackage{makecell} % for line breaks in table cells
\begin{document}

% 图与图注间距（上方）
%\setlength{\abovecaptionskip}{0pt}
% 图注与正文间距（下方）
\setlength{\belowcaptionskip}{0pt}

% \title{SlideFormer: Efficient Full Fine-Tuning on a Single GPU}
% \title{SlideFormer: An Efficient Co-Design for Fine-Tuning on Single GPU}
\title{An Efficient Heterogeneous Co-Design for Fine-Tuning on a Single GPU}
%\title{SlideFormer: A Unified Layer-Sliding Architecture for Heterogeneous Memory Co-Design}

\author{Ruijia Yang}
%\orcid{0009-0005-0022-7840}
\affiliation{%
  \institution{HKUST(GZ)}
  \city{Guangzhou}
  \country{China}
}
\email{ryang379@connect.hkust-gz.edu.cn}

\author{Zeyi Wen}
\authornote{Corresponding author: wenzeyi@hkust-gz.edu.cn}
\affiliation{
  \institution{HKUST(GZ) \& HKUST}
  \city{Guangzhou}
  \country{China}
}
\email{wenzeyi@hkust-gz.edu.cn}

\renewcommand{\shortauthors}{Yang and Wen}

\begin{abstract}
Fine-tuning Large Language Models (LLMs) has become essential for domain adaptation, but its memory-intensive property exceeds the capabilities of most GPUs. To address this challenge and democratize LLM fine-tuning, we present SlideFormer, a novel system designed for single-GPU environments. Our innovations are: (1) A lightweight asynchronous engine that treats the GPU as a sliding window and overlaps GPU computation with CPU updates and multi-tier I/O. (2) A highly efficient heterogeneous memory management scheme significantly reduces peak memory usage. (3) Optimized Triton kernels to solve key bottlenecks and integrated advanced I/O. This collaborative design enables fine-tuning of the latest 123B+ models on a single RTX 4090, supporting up to 8× larger batch sizes and 6× larger models. In evaluations, SlideFormer achieves 1.40× to 6.27× higher throughput while roughly halving CPU/GPU memory usage compared to baselines, sustaining >95\% peak performance on both NVIDIA and AMD GPUs. The code is available at \url{https://github.com/RegiaYoung/SlideFormer}.

\end{abstract}


% \ccsdesc[300]{Computing methodologies~Parallel computing methodologies}

% \keywords{Do, Not, Use, This, Code, Put, the, Correct, Terms, for,


\maketitle

\section{Introduction}
\label{sec:intro}

Large Language Models (LLMs) have revolutionized natural language processing with their remarkable capabilities across diverse tasks~\cite{radford2019language,mann2020language}, and fine-tuning open-source pre-trained models~\cite{grattafiori2024llama3herdmodels,qwen2025qwen25technicalreport,mistralai2024mistral} on specific datasets is often preferred over training from scratch to achieve specialized performance~\cite{wei2021finetuned}. However, as the models continue to grow in size, their fine-tuning memory requirements increase linearly. For example, fine-tuning an 8B model with mixed precision training~\cite{micikevicius2017mixed} requires over 128 GB of GPU memory, far exceeding the VRAM of most high-end GPUs (e.g., 24-96 GB).

%However, the democratization of this technology is hindered by a significant hardware barrier. Full-parameter fine-tuning of a model like Llama-3.1 8B~\cite{grattafiori2024llama3herdmodels} requires over 128GB of memory, far exceeding the VRAM of most high-end GPUs (e.g., 24-96GB).

This memory bottleneck prevents the democratization of LLM fine-tuning, posing a significant barrier for individuals and small labs without access to GPU clusters or cloud resources. For single-GPU scenarios, a paradox arises: modern GPUs such as RTX 4090 possess ample computational power to fine-tune an 8B model, yet existing methods cannot efficiently handle the bottleneck, creating an urgent need for single-GPU solutions that break the VRAM wall.

\begin{figure}[htbp]
    %\vspace{-0.1cm}
    \centering
    \includegraphics[width=\linewidth]{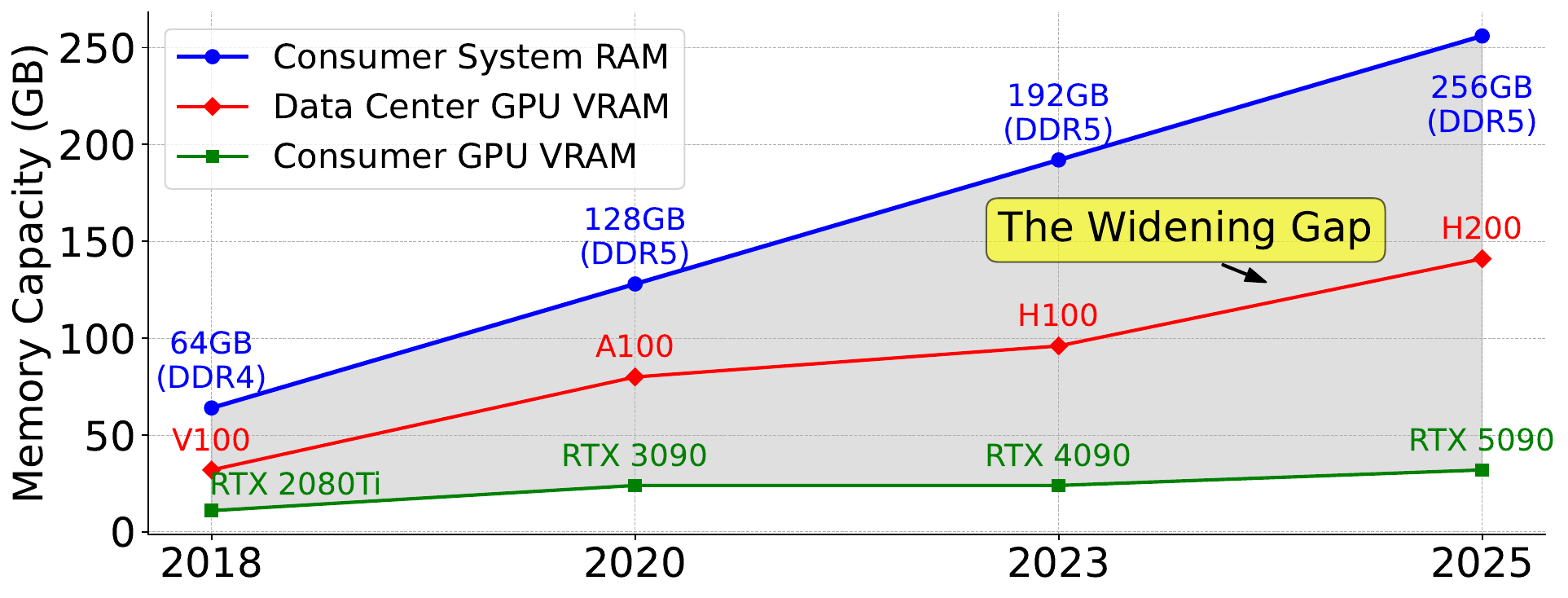}
    \vspace{-0.6cm}
    \caption{The widening gap between CPU and GPU memory.}
    \label{fig:trend} % (见下方关于标签的说明)
\end{figure}

A key trend motivating our work is the increasingly divergent growth trajectories between CPU and GPU memory, as shown in figure~\ref{fig:trend}. Consumer systems now utilize DDR5 memory with doubled capacity (up to 256 GB) and faster I/O (PCIe and NVMe), whereas the maximum VRAM on GPUs has seen modest increases, from 24 GB (RTX 3090) in 2020 to 32 GB (RTX 5090) by 2025. This widening gap makes offloading attractive, turning single-GPU fine-tuning into a heterogeneous system design problem: How can we holistically co-design a system to leverage the entire platform (GPU, CPU, RAM, NVMe) to overcome the VRAM bottleneck?

Various methods have been proposed to address the memory constraints in LLM fine-tuning. Distributed techniques such as Pipeline Parallelism~\cite{huang2019gpipe,narayanan2019pipedream}, Tensor Parallelism~\cite{shoeybi2019megatron}, and Data Parallelism~\cite{li2020pytorch,rajbhandari2020zero} are generally unsuitable for single-GPU scenarios. Parameter-efficient fine-tuning~\cite{peft} methods such as LoRA~\cite{hu2022lora} have been proven insufficient to match the performance of full parameter fine-tuning in many cases~\cite{shuttleworth2025loravsfinetuningillusion}. Among existing offloading systems, ZeRO-Offload~\cite{ren2021zero} and ZeRO-Infinity~\cite{rajbhandari2021zero} are widely recognized. However, their designs are primarily for multi-GPU settings and fail to effectively pipeline computation with transfers and CPU updates, leaving significant room for performance improvement in single-GPU scenarios. Although some works~\cite{sun2022stronghold,jang2024smart,liao2024lohanlowcosthighperformanceframework} have explored this overlap potential, they are incompatible with recent LLMs and lack fine-grained optimizations for memory and efficiency, which are critical for practical usability.

To address the challenge, we present SlideFormer, a novel framework optimized for single-GPU fine-tuning through holistic heterogeneous co-design. Our work makes the following contributions:

%(1) \textbf{Layer-Sliding Mechanism}: We dynamically manage GPU memory by keeping only a small fraction of active layers on the GPU, with parameters and gradients efficiently loaded and offloaded from CPU memory.
%(2) \textbf{Asynchronous Offloading with Multi-Pipelines}: We implemented multi-tiered pipelines with dedicated CUDA streams and CPU threads to overlap GPU computation with data transfers and CPU-bound operations (parameter updates), maximizing the utilization of heterogeneous resources. 
%(3) \textbf{Memory-Efficient Cuda Kernels}: We also integrate optimized CUDA kernels to reduce memory consumption and improve computational efficiency, especially a fused operator for loss computation, which is often overlooked. 
%(4) \textbf{Extended Storage Hierarchy}: We support strategic offloading to NVMe storage with optimized I/O operations like GPUDirect Storage and pipelines, further extending the range of models that can be fine-tuned on limited hardwar

•\: \textbf{A Lightweight Asynchronous Engine:} We propose a Layer-Sliding architecture that maintains a small, active window on the GPU, orchestrated by a multi-pipeline engine built on a lightweight thread-based mechanism, which efficiently overlaps GPU computation with CPU updates and I/O across hierarchies.

•\: \textbf{Efficient Heterogeneous Memory Management:} A queue of pre-allocated GPU cache units eliminates fragmentation and reallocation, while host-side shared buffers for gradients and type conversion reduce peak CPU memory by over 25\%. In concert with our pipeline, this co-design enables fine-tuning with significantly less GPU and CPU memory than prior work.

•\: \textbf{Integrated Advanced I/O and Optimized Kernels:} We extend the memory hierarchy to NVMe and pioneer the integration of GPUDirect Storage\cite{nvidia_gpudirectstorage} for offloading, bypassing the CPU. We also integrate a suite of fused Triton kernels for computations, resolving critical memory bottlenecks overlooked by previous systems.

The holistic co-design translates directly to state-of-the-art performance and scalability, enabling fine-tuning >123B models on a single RTX 4090. For a high-end PC equipped with 256 GB CPU memory, models up to 24B can be fine-tuned at over 95\% peak GPU performance on both NVIDIA and AMD GPUs. Compared to existing frameworks, SlideFormer achieves a 1.40× to 6.27× improvement in throughput, reduces GPU memory consumption by over 50\%, lowers CPU memory usage by approximately 40\%, and supports 8x larger batch sizes and 6× larger model sizes. 

Our work is implemented based on PyTorch~\cite{li2020pytorch} and Transformers~\cite{wolf-etal-2020-transformers} libraries, ensuring compatibility with the latest model architectures (e.g., Llama, Qwen). We expect SlideFormer to democratize LLM fine-tuning, enabling individuals and researchers with limited resources to leverage the power of large models. 

%--------------------------------BG-------------------------------------------
% \begin{table*}[t]
% \small
% \centering
% \caption{Comparison of recent offloading frameworks, highlighting SlideFormer's advanced designs.}
% \label{tab:framework_comparison}
% \begin{tabular}{l|c|c|c|c}
% \hline
% \textbf{Feature} & \textbf{ZeRO-Offload~\cite{ren2021zero}} & \textbf{StrongHold~\cite{sun2022stronghold}} & \textbf{LoHan~\cite{liao2024lohanlowcosthighperformanceframework}} & \textbf{SlideFormer (Ours)} \\
% \hline
% \hline
% \textbf{Backward Overlap} & $\times$ & \checkmark & \checkmark & \checkmark \\
% \hline
% \textbf{Async Mechanism} & Stream & Stream \& ThreadPool & Stream \& Multiprocess & \textbf{Stream \& ThreadPool} \\
% \hline
% \textbf{Operation Granularity} & Param Group & Layer & Param Group & \textbf{Layer} \\
% \hline
% \textbf{Memory Management} & On-demand & On-demand & On-demand & \textbf{Pre-allocated} \\
% \hline
% \textbf{Activation Offloading} & $\times$ & CPU & CPU/NVMe &  \textbf{CPU/NVMe} \\
% \hline
% \textbf{NVMe Offload Techs} & DeepSpeed AIO & PyTorch Serialization & DeepSpeed AIO & \textbf{GPUDirect Storage} \\
% \hline
% \textbf{Optimized Kernels} & $\times$ & Megatron Kernels & $\times$ & \textbf{Custom Kernels} \\
% \hline
% \textbf{LLM Support} & Broad & Limited  & Limited & \textbf{Broad} \\
% \hline
% \end{tabular}%
% \end{table*}

\section{Background}
\label{sec:background}

\subsection{Memory Challenges in LLM Fine-Tuning}
\label{sec:mem-challenges}
Fine-tuning adapts a pre-trained LLM to a target domain with far fewer steps and data than pre-training; yet it remains memory-bound at scale. For a model with $N$ parameters and $n$ layers, hidden size $h$, sequence length $s$, and batch size $b$, memory demand comes from: parameters, gradients, optimizer states, and activations.

\textbf{Static footprints.}
Parameters are typically stored in \texttt{FP16/BF16} ($2N$ bytes), while gradients contribute another $2N$ in \texttt{FP16/BF16}. The Adam~\cite{kingma2014adam} optimizer is commonly used; it adds two \texttt{FP32} states per parameter (momentum/variance, $8N$), making optimizer states the largest static term. Besides, mixed-precision training~\cite{micikevicius2017mixed} requires the optimizer to maintain an \texttt{FP32} master copy ($4N$) of parameters for stability. Forward activations scale with $\mathcal{O}(n\!\cdot\!h\!\cdot\!s\!\cdot\!b)$ and must be available for the backward pass unless recomputed. A succinct approximation is:
\begin{equation}
\label{mem_req} %\setlength{\abovedisplayskip}{0pt}
Mem_{req} = \underbrace{2N}_{\text{Params}} + \underbrace{2N}_{\text{Grads}} + \underbrace{4N \: + \: 8N}_{\text{Optimizer States}} + \underbrace{\mathcal{O}(n \cdot h \cdot s \cdot b)}_{\text{Activations}}
\end{equation}

\textbf{Single-GPU tension.}
Distributed parallelism techniques~\cite{huang2019gpipe,narayanan2019pipedream,shoeybi2019megatron,li2020pytorch,rajbhandari2020zero} amortize memory across multiple devices but are infeasible on a single GPU. A single high-end GPU has ample compute to fine-tune multi-billion-parameter models; yet the footprint in Eq.~\eqref{mem_req} frequently exceeds VRAM, forming the central bottleneck.

\textbf{Common mitigations.}
\emph{Gradient checkpointing}~\cite{chen2016training} trades 30\% extra compute for $>\!80\%$ activation savings; \emph{PEFT} (e.g., Adapter~\cite{houlsby2019parameter}, LoRA~\cite{hu2022lora}) updates a small subset of weights but underperforms compared to full-parameter fine-tuning on domain-critical tasks~\cite{wei2021finetuned,zhao2024galore,NEURIPS2024_2c570b0f,peft}; \emph{Kernel optimizations} (e.g., FlashAttention~\cite{dao2022flashattention}, xFormers~\cite{xFormers2022}, Liger~\cite{hsu2024ligerkernelefficienttriton}) reduce transient allocations and improve throughput. These techniques are complementary, but not sufficient to resolve the VRAM wall in single-GPU full-parameter fine-tuning.

\subsection{Existing Offloading Techniques}

A key trend driving us is the increasingly divergent growth trajectories between CPU and GPU memory, as shown in Figure~\ref{fig:trend}. Recent PCs and workstations with abundant CPU memory (e.g., up to 256 GB DDR5) and high-speed NVMe storage enable memory-efficient LLM fine-tuning through strategic offloading. Coupled with faster PCIe interconnects, stronger CPU performance, and technologies like GPUDirect Storage~\cite{nvidia_gpudirectstorage}, this motivates a pipeline-aware offloading design that jointly orchestrates the GPU, CPU, and NVMe rather than treating VRAM as the only limiting resource.

Several representative frameworks have been developed to this end. ZeRO-Offload~\cite{ren2021zero} pioneers offloading the optimizer and gradients to the CPU. Then ZeRO-Infinity~\cite{rajbhandari2021zero} extends it to a multi-tiered memory system, dynamically offloading components to both CPU and NVMe. Other notable systems, such as Transformer Engine~\cite{TransformerEngine2024} and NeMo~\cite{NVIDIANeMo}, provide a layer-wise approach for activation offloading, and ColossalAI~\cite{ColossalAI}'s \textit{gemini}~\cite{colossalai_gemini} introduces a dynamic chunk-based hetero-memory management. Besides, several research prototypes have explored similar concepts~\cite{sun2022stronghold,jang2024smart,liao2024lohanlowcosthighperformanceframework}. 

% 现有的主流框架主要还是面向并行和分布式场景，应用于单GPU场景有诸多限制。具体而言，Zero-Offload和Zero-Infinity虽然在多卡微调上相当有效，但其设计并不是针对单 GPU 场景专门设计的，并不能高效地重叠计算和传输，同时显存占用也高;虽然Transformer Engine提出了layer-granular的卸载，但只用于激活没有进一步地扩展; 同样的, ColossalAI的chunk-wise的细粒度内存管理也是面向分布式场景的，虽然能比较好地利用CPU和GPU内存而放入更大的模型, 但在单GPU上的效率很低。Stronghold~\cite{sun2022stronghold}首次提出了将后向传播和参数更新重叠的关键见解，但其在旧版本Megatron上修改的实现已经outdated，当时其并未意识到这个设计在单GPU场景上的潜力，也未做相应优化，我们的设计承袭这一设计如Figure~\ref{},并针对场景一步步分析并优化性能和技术细节~\ref{}，成为单GPU场景下最优的可用方法；有其他工作也注意到了类似设计的潜力~\cite{wang2025zo2scalablezerothorderfinetuning,liao2024lohanlowcosthighperformanceframework}, 有的迁移至非传统的优化器，有的基于更细粒度的卸载，但他们的实现不是完整框架可实际使用级别的，同时基于很老旧的模型结构，缺乏对常用的（llama, mistral, qwen）的支持，并没有对single-GPU这个场景所要面对的challenge分析和优化。

\subsection{Limitations of Existing Solutions}
\label{sec:limitations}

While mainstream frameworks excel in distributed and multi-GPU settings, their design is not holistically co-designed for single-GPU scenarios. For instance, ZeRO-Offload and ZeRO-Infinity inherit considerable overhead from their distributed-first architecture; mechanisms intended for multi-GPU communication remain active on a single device, introducing additional memory footprint and latency. This, combined with underutilized CPU memory pools, creates significant overhead, as observed in Section~\ref{sec:heter_mem}. Similarly, ColossalAI's chunk-wise memory management, while effectively utilizing memory for larger models, is suboptimal for single-GPU efficiency. Critically, their design is synchronous at the update stage, leaving the GPU idle while waiting for the CPU update to finish.

Academic prototypes that recognize this overlap potential still suffer from critical design flaws. Stronghold~\cite{sun2022stronghold} was an early attempt but relied on an outdated version of Megatron~\cite{shoeybi2019megatron} and did not fully recognize or optimize for single-GPU environments. 
LoHan~\cite{liao2024lohanlowcosthighperformanceframework}, a recent work, employs a multiprocess-based engine for asynchronous updates, which incurs IPC overhead, rather than a thread-based approach. Furthermore, LoHan utilizes on-demand memory management, which is prone to runtime fragmentation, and operates at a Param Group granularity without analyzing how to set its size. Its design choices are architecturally distinct from SlideFormer's pre-allocated and layer-granular design. These limitations, combined with incomplete optimizations (e.g., ignoring the CrossEntropyLoss bottleneck) and limited model support (e.g., only GPT-2), necessitate a new, holistically designed system.

\section{System Design}
\label{sec:design}

\begin{figure}[h]
    \centering
    \includegraphics[width=\linewidth]{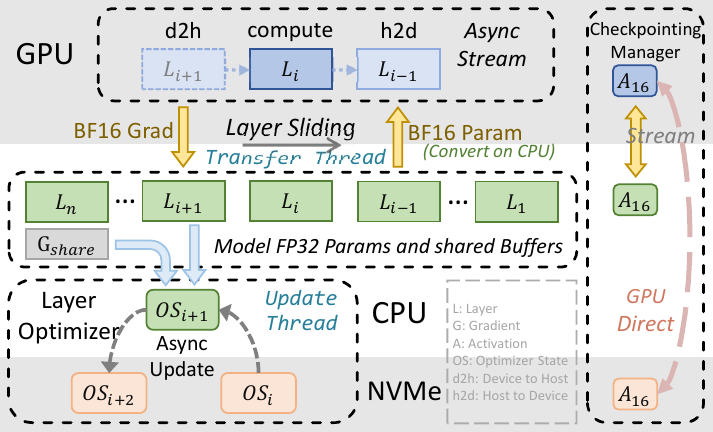}
    \vspace{-0.65cm}
    \caption{Overview of SlideFormer.}
    \label{fig:overview}
\end{figure}

The design goal of SlideFormer is to break the memory wall of single-GPU fine-tuning through a holistic system-level co-design, while achieving state-of-the-art efficiency. We propose a unified architecture where computation scheduling, memory management, and I/O are jointly optimized. As illustrated in Figure~\ref{fig:overview}, our system is built on three pillars: 
(1) a Layer-Sliding Architecture powered by a lightweight asynchronous engine, 
(2) a Pre-allocated Heterogeneous Memory system to eliminate overhead, 
(3) an Integrated I/O and Compute stack utilizing GPUDirect and fused kernels.

\subsection{The Layer-Sliding Architecture}
\label{sec:async-offloading}
\begin{figure}[htbp]
    \centering
    \includegraphics[width=\linewidth]{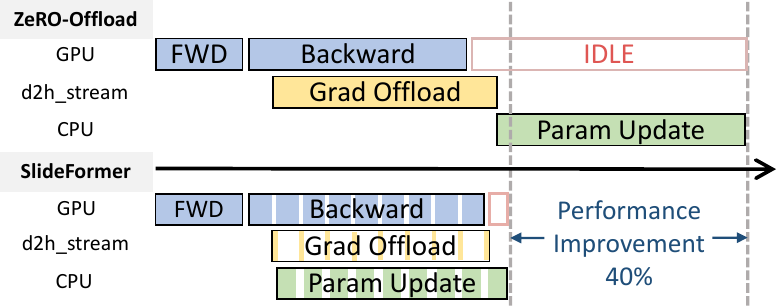}
    \vspace{-0.6cm}
    \caption{Backward overlaps with parameter updates.}
    \label{fig:backward}
\end{figure}

\textbf{Asynchronous Parameter Updating:}
As illustrated in Figure~\ref{fig:backward}, we adopt a layer-granular approach to pipeline the backward and update with offloading. Once the backward computation for layer $L_{i}$ finishes on the GPU, its gradients $G_i$ are asynchronously transferred to host memory (d2h). In parallel, the CPU applies the optimizer to update $P_i$ using the host-resident optimizer states. While the CPU updates $P_i$, the GPU continues computing the backward pass for $L_{i-1}$ and prefetches the parameters for $L_{i-2}$ (h2d). This schedule eliminates the heterogeneous resource idle issue in ZeRO-Offload~\cite{ren2021zero} by overlapping GPU-bound compute with CPU-bound updates and cross-tier transfers.

\begin{figure}[!htbp]
    \centering
    \includegraphics[width=0.98\columnwidth]{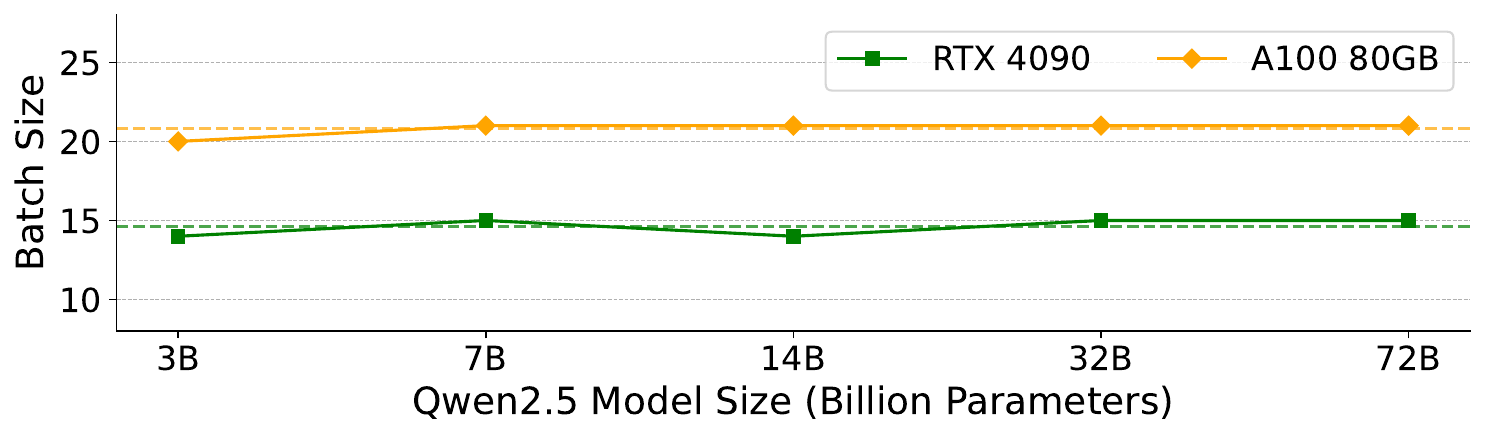}
    \vspace{-0.4cm}
    \caption{
        Critical batch size for achieving full backward overlap with updates ($T_{bwd} \geq T_{grad\_d2h} + T_{update}$).
    }
    \label{fig:critical_batch_size}
\end{figure}

\textbf{Rationale for Layer Granularity:} The cornerstone of efficiency lies in our layer-granular strategy for memory management and computation scheduling, which restructures the fine-tuning process to maximize Hetero-hardware utilization. Layer is the smallest constitutional repeating unit in LLMs. Non-repeating units, such as the param-group used in ZeRO-Offload or LoHan~\cite{liao2024lohanlowcosthighperformanceframework}, introduce complex management for various-sized components and require manual configuration. Critically, a multi-layer window is counterproductive in memory-constrained environments, as layers are computed serially, consuming scarce VRAM that could be used to increase the batch/model size while offering negligible benefits. As shown in Figure~\ref{fig:critical_batch_size}, the critical batch size required to achieve effective overlap remains remarkably stable across different layer sizes (from 77M-3B to 878M-72B). Because all backward pipeline latencies ($T_{bwd}$, $T_{grad\_d2h}$, $T_{update}$) scale proportionally with granularity, the overlap condition mainly depends on the batch size. A single layer is sufficient to saturate modern GPU, as evidenced by the high GPU utilization in Table~\ref{tab:pipeline_overlap} and Figure~\ref{fig:qwen_comparison}.

\textbf{Thread-Based Lightweight Engine:} 
The backbone of SlideFormer's efficiency is its extensive use of asynchronous operations to overlap data transfers and CPU computations with the GPU workload. Unlike LoHan, which relies on a multi-process optimizer introducing IPC overhead, SlideFormer implements a lightweight thread-based engine through dedicated:
(i) \textit{CUDA Streams~\cite{nvidia_cuda_stream}}: Separate streams are employed for asynchronous h2d/d2h transfers and concurrent GPU computation.
(ii) \textit{CPU Threads}: Two thread executors, one for transfers between h2d/d2h and the other for Layer-Adam to update parameters, prevent potential blocking I/O or CPU-intensive tasks from stalling the main fine-tuning thread. 

\begin{figure}[htbp]
    \centering
    \includegraphics[width=\linewidth]{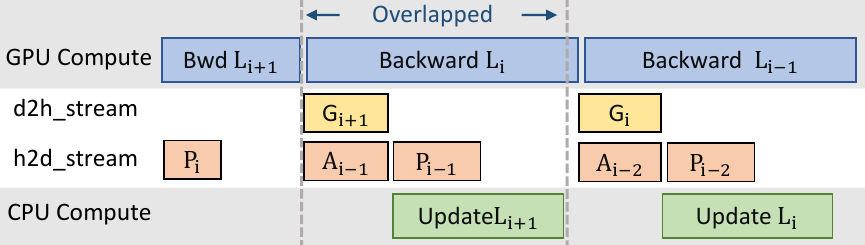}
    \vspace{-0.6cm}
    \caption{Computation-communication overlap during backward propagation in GPU-CPU tier pipeline.}
    \label{fig:pipeline}
\end{figure}

\textbf{Condition for Effective Overlap:}
The efficiency of our asynchronous engine hinges on latency hiding, where the following conditions should be met: (i) In forward pass, lossless overlap occurs when the computation time for the current layer is greater than or equal to the parameter prefetch time for the next layer, i.e., $T_{compute\_fwd} \geq T_{param\_h2d}$. (ii) In the backward pass, as illustrated in Figure~\ref{fig:pipeline}, lossless overlap occurs when $T_{compute\_bwd} \geq T_{grad\_d2h} + T_{update}$. When NVMe offloading is enabled, the transfer overhead of the optimizer states makes $T_{update}$ the main performance bottleneck. To quantify the degree of backward overlap, we introduce the \textbf{hiding factor} ($\eta = T_{\text{bwd}} / (T_{\text{d2h}} + T_{\text{update}})$), where $\eta\geq 1$ indicates zero-overhead offloading. Table~\ref{tab:pipeline_overlap} presents the timeline breakdown for fine-tuning Qwen2.5-14B, confirming that our architecture achieves effective overlap across various hardware. Unlike sequential methods such as ZeRO-Offload that would completely stall the GPU, SlideFormer maintains a robust performance advantage even on imbalanced hardware where full overlap ($\eta<1$) is infeasible, using extremely powerful or memory-limited GPUs.

% \begin{table}[!ht]
% \centering
% \small
% \caption{
%     Profile timelines of backward stage for SlideFormer during the fine-tuning of Qwen2.5-14B. (All time in ms)
% }
% \label{tab:pipeline_overlap}
% \small
% \begin{tabular}{cccccc}
% \toprule

% \textbf{Batch Size} & \textbf{$T_{bwd}$} & \textbf{$T_{d2h}$} & \textbf{$T_{update}$} & \textbf{Factor} ($\eta$) & \textbf{GPU Util. (\%)} \\
% \midrule
% \multicolumn{6}{c}{\textit{RTX 4090 24GB (PC)}} \\
% \midrule
% 16 & 170 & 22 & 175 & \textbf{0.66} & 93.1 \\
% 32 & 340 & 25 & 195 & \textbf{1.55} & 96.9 \\
% 64 & 660 & 25 & 195 & \textbf{3.00} & 98.4 \\
% \midrule
% \multicolumn{6}{c}{\textit{A100 80GB (Server)}} \\
% \midrule
% 32 & 225 & 24 & 152 & \textbf{1.28} & 97.2 \\
% 64 & 450 & 25 & 151 & \textbf{2.56} & 98.8 \\
% 128 & 910 & 25 & 153 & \textbf{5.11} & 99.3 \\
% \bottomrule
% \end{tabular}
% \end{table}

\begin{table}[!ht]
\centering
\small
\caption{Profile timelines of backward stage for SlideFormer during the fine-tuning of Qwen2.5-14B. (All time in ms)}
\label{tab:pipeline_overlap}
\vspace{-0.2cm}
\setlength{\tabcolsep}{3.5pt}
\renewcommand{\arraystretch}{1.05}

\begin{tabular}{lcccccc}
\toprule
\textbf{System} & \makecell{\textbf{Batch}\\\textbf{Size}} & \textbf{$T_{bwd}$} & \textbf{$T_{d2h}$} & \textbf{$T_{update}$} & \makecell{\textbf{Factor}\\($\eta$)} & \makecell{\textbf{GPU Util.}\\\textbf{(\%)}} \\
\midrule
\multirow{3}{*}{\makecell[l]{\textit{RTX4090 24GB}\\\textit{(PC)}}} 
 & 16  & 170 & 22 & 175 & \textbf{0.66} & 93.1 \\
 & 32  & 340 & 25 & 195 & \textbf{1.55} & 96.9 \\
 & 64  & 660 & 25 & 195 & \textbf{3.00} & 98.4 \\
\midrule
\multirow{3}{*}{\makecell[l]{\textit{A100 80GB}\\\textit{(Server)}}} 
 & 32  & 225 & 24 & 152 & \textbf{1.28} & 97.2 \\
 & 64  & 450 & 25 & 151 & \textbf{2.56} & 98.8 \\
 & 128 & 910 & 25 & 153 & \textbf{5.11} & 99.3 \\
\bottomrule
\end{tabular}
\end{table}

\subsection{Efficient Heterogeneous Memory Co-Design}
\label{sec:heter_mem}

Previous works~\cite{ren2021zero,sun2022stronghold,liao2024lohanlowcosthighperformanceframework} often overlooked the evaluation and optimization of heterogeneous memory footprints, but we hope to co-design an extremely efficient, fixed footprint, and fragment free memory management system based on a layer sliding architecture.

\textbf{Pre-allocated GPU Cache Unit Queue:} Rather than keeping the entire model in GPU memory, SlideFormer maintains a window of active layers, which is exactly a queue of pre-allocated GPU cache units, each sized to hold a layer's parameters and gradients. During training, layers (i.e., parameters) sequentially slide into this cache queue to perform computations, after which the used units are released for new layers. Only during the backward pass, the gradients of each layer are offloaded to CPU memory. Unlike the \textit{on-demand} allocation used by StrongHold~\cite{sun2022stronghold} and LoHan~\cite{liao2024lohanlowcosthighperformanceframework}, this unit reuse design ensures a fixed GPU memory footprint and avoids reallocation, reducing overhead and fragmentation.

\textbf{Optimized CPU Memory Layout with Shared Buffers:} On the CPU side, \texttt{FP32} parameter master copies of each layer are stored in a flattened, pinned tensor (\texttt{cpu\_params\_flat}) for efficient h2d transfers. To optimize memory usage, we employ shared buffers for intermediate data. Gradients offloaded from the GPU are stored in a layer-shared, pinned \texttt{BF16/FP16} tensor (\texttt{cpu\_grad\_flat}), which reduces the gradient footprint on CPU memory ($2N$ bytes) to \(1 / num\_layers\). Similarly, a layer-shared buffer is dedicated to convert \texttt{FP32} parameters to \texttt{BF16/FP16} before h2d transfer, thus avoiding additional transfer/memory costs of type conversion on the GPU and storing $2N$ bytes of \texttt{BF16/FP16} parameters in CPU memory. On the GPU side, parameters and gradients maintain \texttt{BF16/FP16} precision, following the mixed precision training~\cite{micikevicius2017mixed} scheme.

\textbf{Sliding Activation}: To further alleviate GPU memory pressure from activations, we employ a sliding checkpointing mechanism modified from standard gradient checkpointing~\cite{chen2016training, li2020pytorch}. After each layer's forward pass, activations are asynchronously offloaded to the CPU memory or NVMe and prefetched to the GPU memory for recomputation before the backward pass of that layer, ensuring that VRAM required for activations is limited to only a small window. We pre-allocate pinned tensors in CPU memory or files on SSDs for storing activations before the fine-tuning begins.

\textbf{Layer-Adam Optimizer}: A self-developed variant of DeepSpeed's CPU-Adam, it stores the optimizer states of each layer in a flattened tensor in the host memory. When the gradients of the layer are offloaded to the CPU, the optimizer updates the layer's parameters separately. Additionally, the optimizer states can be further offloaded to the NVMe tier, and an asynchronous offload-prefetch mechanism is established to reduce latency. 

\subsection{Integrated I/O and Compute Co-Design}
\label{sec:io_compute}
The final pillar of our co-design optimizes the data movement paths and intra-layer computation to eliminate remaining bottlenecks that pure scheduling cannot address.

\textbf{GPUDirect Storage and NVMe Tiering:} To support models exceeding CPU RAM capacity, SlideFormer extends the memory hierarchy to NVMe storage. Crucially, we pioneer to integrate GPUDirect Storage (GDS)~\cite{nvidia_gpudirectstorage} for LLM fine-tuning offload. GDS establishes a direct data path between NVMe and GPU, bypassing the CPU bounce buffer. This "zero-copy" mechanism significantly reduces CPU utilization and PCIe bus contention, leaving CPU resources for asynchronous engine and parameter updates. We support offloading \textit{activations} and \textit{optimizer states} to this NVMe tier.

\textbf{Why Not Offload Parameters.} Although offloading parameters to NVMe storage could achieve lower memory usage and larger models, we deliberately avoid it due to diminishing returns: 
(i) Performance Degradation: Parameter transfers (h2d/d2h) are critical for overlapping with GPU computation (c.f. Section~\ref{sec:async-offloading}). Moving parameters to NVMe would shift the transfer bottleneck from PCIe to NVMe speed, severely hindering overall throughput.
(ii) Simplified Data Paths: As shown in Figure~\ref{fig:overview}, SlideFormer ensures that any given data type moves only  between two memory tiers. Introducing NVMe as a third tier for parameters would complicate the data transfer path and add unnecessary overhead.

\begin{figure}[htbp]
    \centering
    \includegraphics[width=\linewidth]{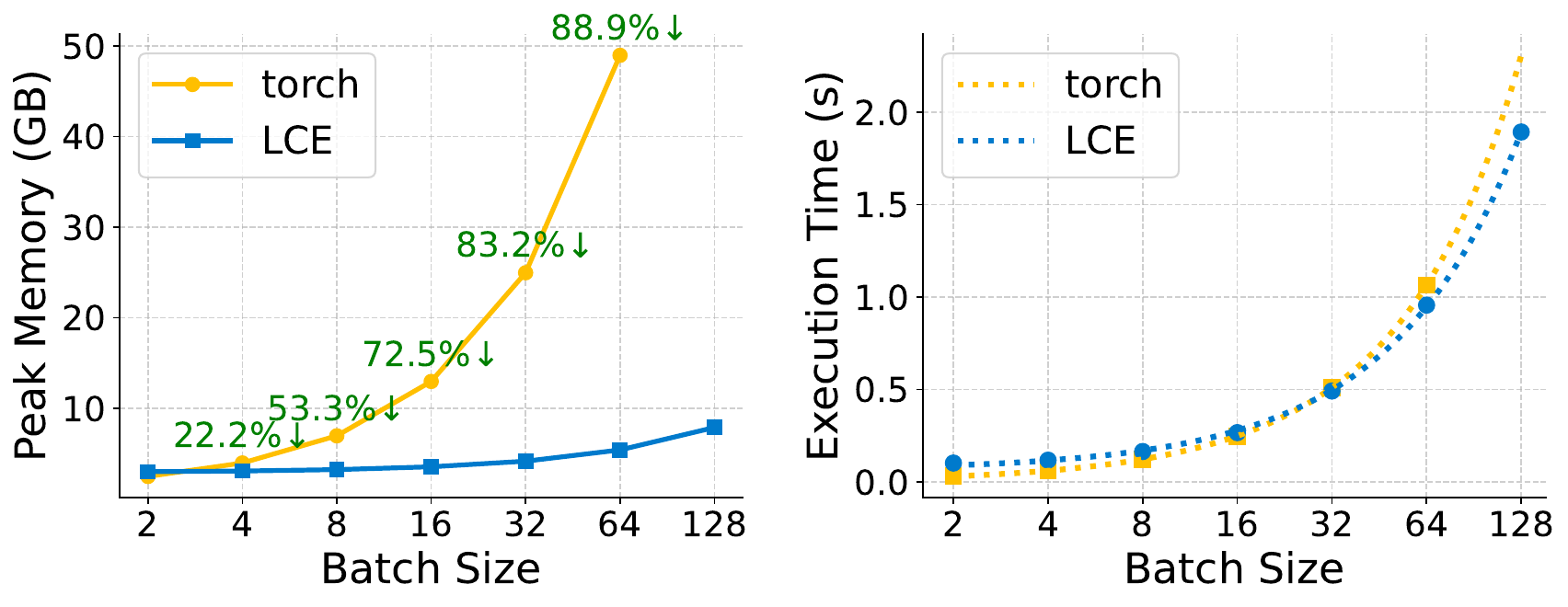}
    \vspace{-0.7cm}
    \caption{Memory usage and execution time comparison between torch standard method and LCE for Llama3.1-8B.}
    \label{fig:lce}
\end{figure}

\begin{figure*}[t]
    \centering
    \begin{minipage}{0.372\textwidth}
        \centering
        % 第一组图片 - Llama（垂直排列）
        \includegraphics[width=\linewidth]{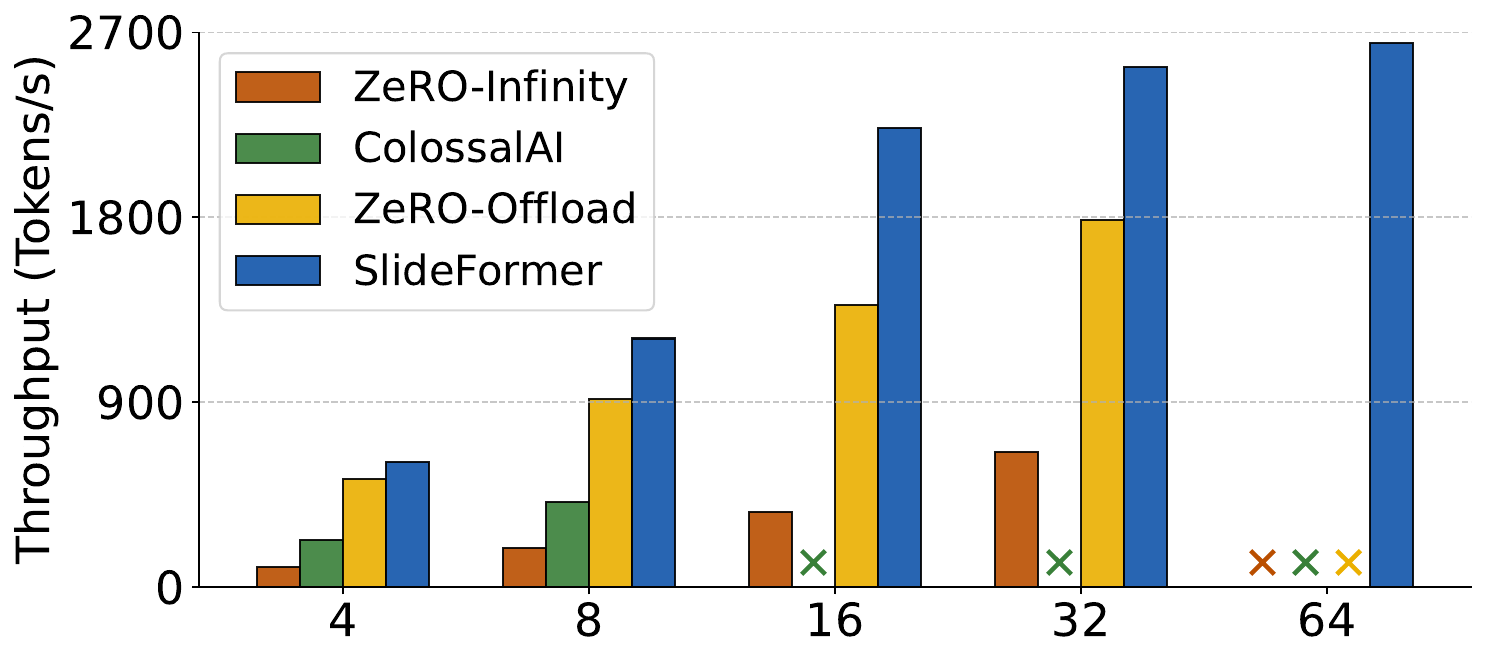}
        % \vspace{0.3cm}
        \includegraphics[width=\linewidth]{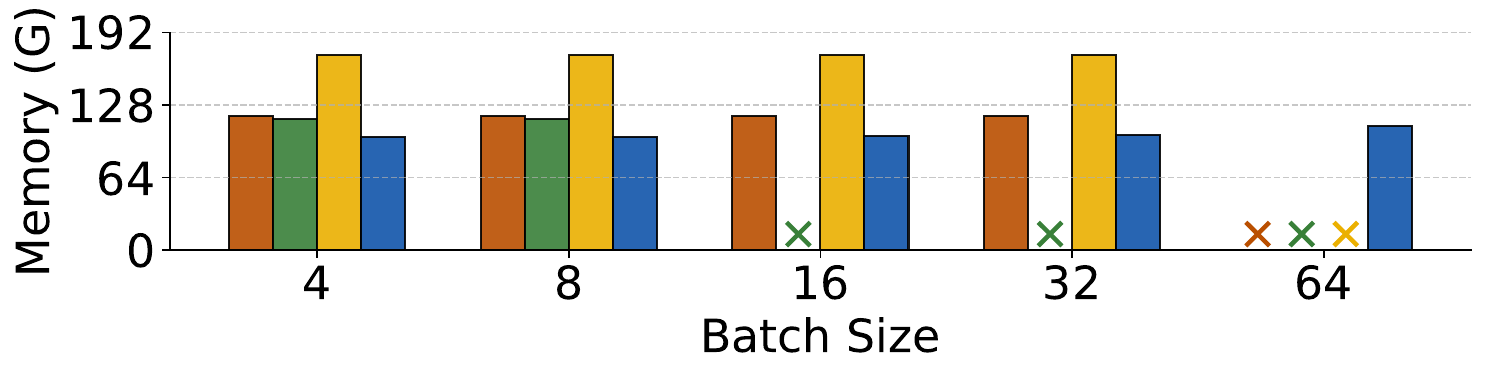}
        \vspace{-0.8cm}
        \caption{Throughput and CPU memory comparison between SlideFormer and baselines for Llama-3.1-8B fine-tuning on RTX4090.}
        \label{fig:llama_comparison}
    \end{minipage}
    \hfill
    \begin{minipage}{0.372\textwidth}
        \centering
        % 第二组图片 - Qwen（垂直排列）
        \includegraphics[width=\linewidth]{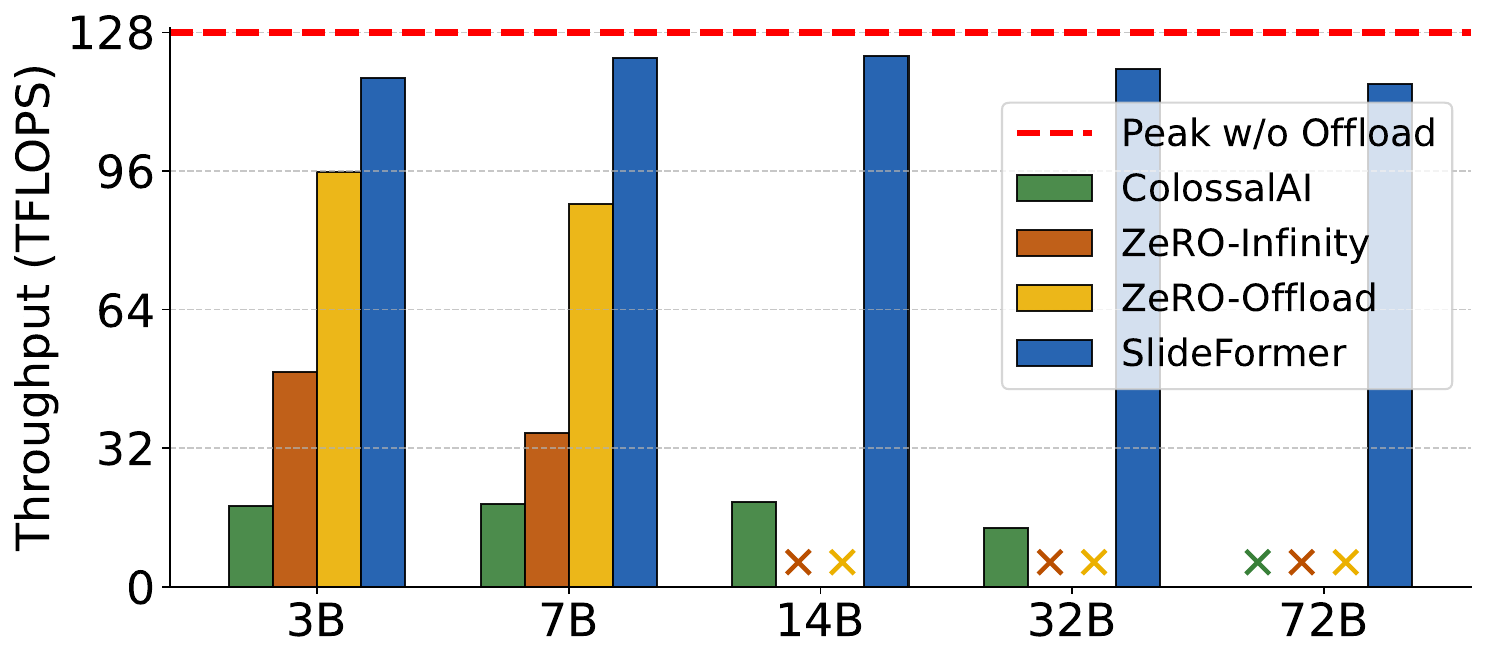}
        % \vspace{0.3cm}
        \includegraphics[width=\linewidth]{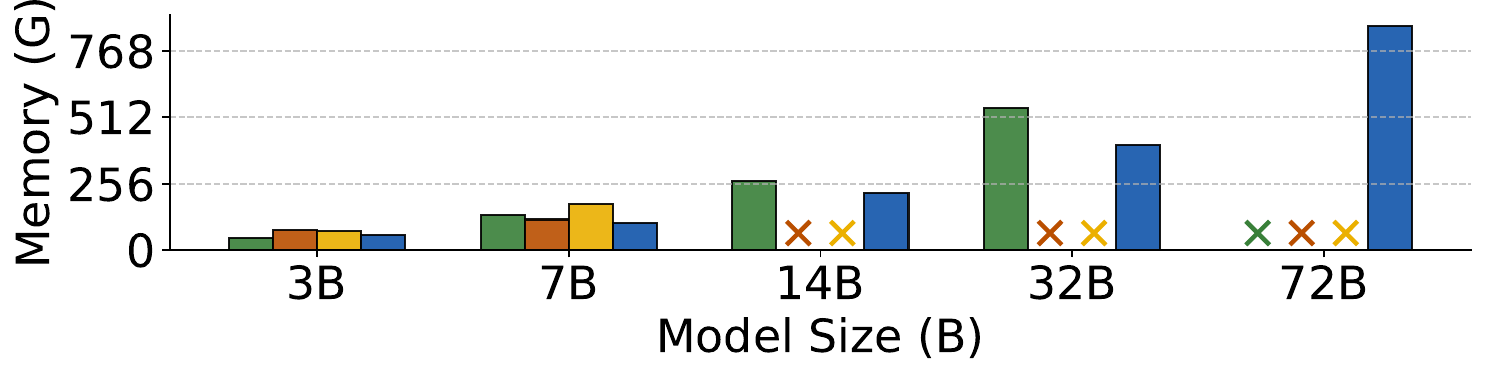}
        \vspace{-0.8cm}
        \caption{Throughput and CPU memory comparison between SlideFormer and baselines for various sizes of Qwen2.5 on RTX4090.}
        \label{fig:qwen_comparison}
    \end{minipage}
    \hfill
    \begin{minipage}{0.244\textwidth}
        \centering
        % 第三组图片 - Mistral（垂直排列）
        \includegraphics[width=\linewidth]{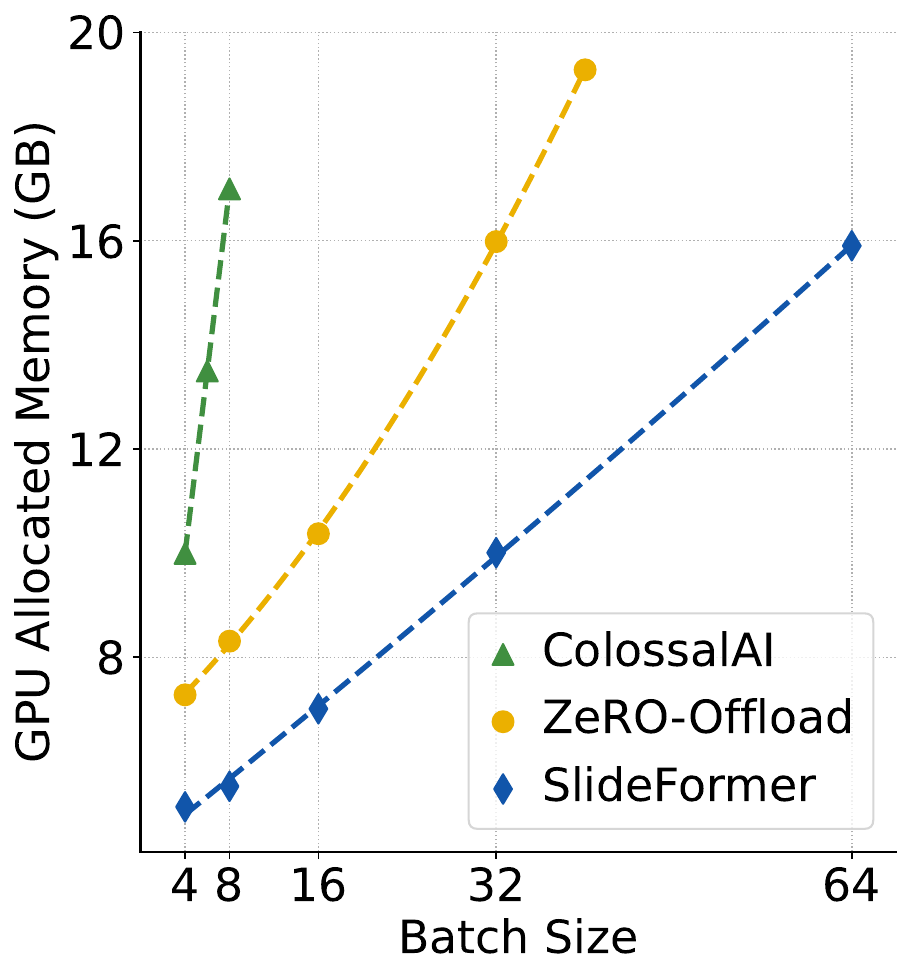}
        \vspace{-0.8cm}
        \caption{GPU memory vs. batch size on various frameworks for Llama-3.1-8B.}
        \label{fig:gpumem}
    \end{minipage}
\end{figure*}

\textbf{Optimized Triton Kernels:} While our pipelines optimize \textit{inter-layer} data movement, we integrate optimized Triton~\cite{triton} kernels to accelerate \textit{intra-layer} computational efficiency. Beyond FlashAttention~\cite{dao2022flashattention}, we employ efficient Triton kernels for operations like RoPE, RMSNorm, and SwiGLU, collectively reducing peak memory usage and improving throughput. Among these, the most critical optimization is the \textbf{fused LinearCrossEntropy kernel} for the output layer and loss computation, which addresses a major and often overlooked memory bottleneck. For recent models with large vocabularies like Llama-3.1, the intermediate logits tensor ($B \times S \times V$) can consume more VRAM than all preceding activations combined. LoHan~\cite{liao2024lohanlowcosthighperformanceframework} sidesteps this issue in evaluation by replacing the standard loss with MSE, which is impractical for real-world tasks. SlideFormer solves this directly by integrating a Fused LinearCrossEntropy (LCE) kernel. This kernel fuses the projection and loss calculation, computing gradients in small chunks to avoid materializing the full logits tensor. As shown in Figure~\ref{fig:lce}, this reduces the memory footprint of the output layer by over 80\% without sacrificing accuracy or speed, unlocking the ability to train with models and batch sizes essential for pipeline saturation.

%--------------------------------------EVAL-------------------------------------

\section{Evaluation}
\label{sec:eval}

In this section, we conduct a comprehensive evaluation of our design to demonstrate its performance and efficiency. 

\subsection{Experimental Setup}

We evaluate SlideFormer on two types of platforms: a high-end PC (NVIDIA RTX 4090 24GB or AMD RX 7900XT 20GB, AMD Ryzen 9 9950X, 256GB DDR5) and a server (NVIDIA A100 80GB, dual Intel Xeon Gold 6338N, 1024GB DDR4). All experiments use PyTorch 2.7.0 and CUDA 12.5 with a fixed sequence length of 1024. For performance benchmarking, we use a synthetic dataset to ensure a consistent computational load (with a stable effective length). 

We compare SlideFormer against leading offloading baselines: ZeRO-Offload~\cite{ren2021zero}, ZeRO-Infinity~\cite{rajbhandari2021zero}, ColossalAI~\cite{ColossalAI}, and LoHan~\cite{liao2024lohanlowcosthighperformanceframework}. To ensure a fair comparison, all frameworks use the latest versions with identical training configs, including activation checkpointing and optimized kernels where applicable. We evaluate a range of modern LLMs, including Llama-3.1 (8B)~\cite{grattafiori2024llama3herdmodels}, Qwen-2.5 (3B-72B)~\cite{qwen2025qwen25technicalreport}, and Mistral (24B-123B)~\cite{mistralai2024mistral}. Performance is measured by Throughput (tokens/s and TFLOPS), peak Memory Usage (GPU and CPU), and trainable model size (B).

\begin{figure}[htbp]
    %\vspace{-0.4cm}
    \centering
    \includegraphics[width=\linewidth]{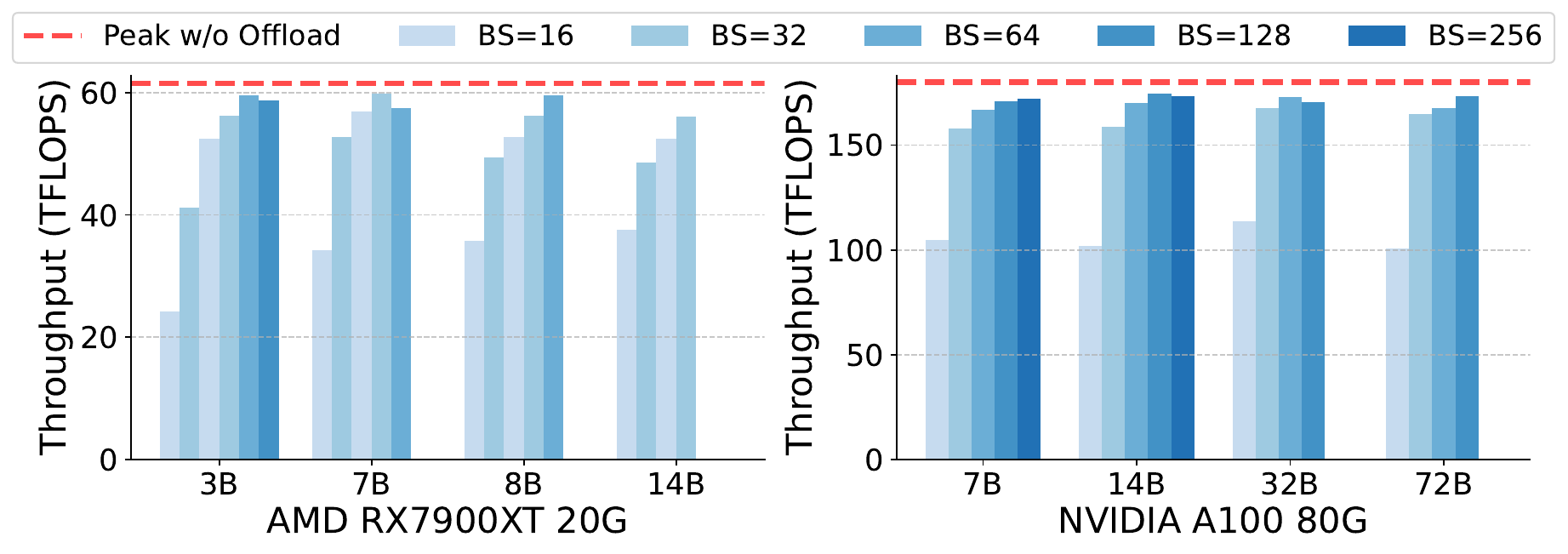}
    \vspace{-0.6cm}
    \caption{The fine-tuning throughput of Qwen2.5 in various sizes on AMD RX7900XT and NVIDIA A100.}
    \label{fig:7900xt} % (见下方关于标签的说明)
\end{figure}

\subsection{Throughput Scalability}
SlideFormer demonstrates superior throughput scalability across both increasing batch sizes and model sizes, consistently outperforming leading offloading systems. 

\textbf{Scalability with Batch Size.} As shown in Figure~\ref{fig:llama_comparison}, SlideFormer outperforms all baselines across every batch size, achieving throughput improvements of 1.39×, 2.82×, 6.34× over baselines on Llama-3.1-8B. The results also illustrate our pipeline's dynamics: at smaller batch sizes, the step time remains constant, as the backward computation is insufficient to fully mask the update latency. However, as the batch size increases to 32, the system shifts to a compute-bound regime where the transfer and update latencies are effectively hidden. This, along with Figure~\ref{fig:7900xt}, confirms our design's ability to leverage larger batch sizes for higher computational throughput.

\textbf{Scalability with Model Size.}
Figure~\ref{fig:qwen_comparison} show that SlideFormer not only delivers higher throughput than baselines at equivalent sizes but also dramatically extends the boundaries of trainable models on a single GPU. While ZeRO-Offload and ZeRO-Infinity fail to run models of 14B parameters or larger, SlideFormer successfully fine-tunes models exceeding 72B parameters. Crucially, SlideFormer's performance consistently reaches 90\% to 95\% of the peak non-offloading fine-tuning TFLOPS. 
This high utilization is robust across platforms, with Figure~\ref{fig:7900xt} confirming similar high efficiency (over 95\% peak performance) on both AMD RX7900XT and NVIDIA A100 GPUs, underscoring SlideFormer's broad applicability.

% \begin{figure*}[t]
%     \centering
%     \begin{minipage}{0.66\textwidth}
%         %\vspace{-0.1cm}
%         \centering
%         \includegraphics[width=\linewidth]{figure/combined_memory.pdf}
%         % \caption{Performance comparison of different offloading strategies}
%         % \label{fig:offloading_comparison}
%         \includegraphics[width=\linewidth]{figure/nvme_count_performance.pdf}
%         \vspace{-0.6cm}
%         \caption{Performance comparison of different NVMe SSD count}
%         \label{fig:nvme_count_comparison}
%     \end{minipage}
%     \hfill
%     \begin{minipage}{0.336\textwidth}
%         % \vspace{+0.2cm}
%         \centering
%         \includegraphics[width=\linewidth]{figure/model_size_memory_comparison.pdf}
%         \vspace{-0.6cm}
%         \caption{Maximum trainable model size}
%         \label{fig:max_model_size}
%     \end{minipage}
% \end{figure*}

\subsection{Heterogeneous Memory Usage}
\label{sec:heter_mem}

SlideFormer's efficient control over memory across the hierarchy is what enables maximum scalability and batch sizes.

\textbf{CPU Memory Efficiency.} The lower panels of Figure~\ref{fig:llama_comparison} and Figure~\ref{fig:qwen_comparison} illustrate that SlideFormer maintains the lowest CPU memory footprint across all scenarios, reducing usage by approximately 40\% compared to the fastest baseline. This significant saving is a direct result of our optimized host memory layout, which utilizes layer-shared buffers for gradients and type conversion, eliminating redundant memory copies and peak consumption.

\textbf{GPU Memory Efficiency.} Figure~\ref{fig:gpumem} plots the GPU memory footprint against batch size, showing that SlideFormer consistently uses the least VRAM, achieving a reduction of over 50\% compared to ZeRO-Offload. This is attributed to our pre-allocated cache queue and the integrated Fused LCE kernel, which together alleviate the primary memory bottleneck in fine-tuning, making it feasible to train large models on consumer-grade hardware.

For example, an individual with a PC with 128GB CPU memory can fine-tune the Llama-3.1-8B model on a single RTX 4080 GPU. This is achievable on a single GPU without resorting to NVMe offloading, while maintaining nearly lossless throughput compared to non-offloaded training. This capability is a cornerstone of our goal to democratize access to large model fine-tuning.

\begin{figure}[htbp]
    \centering
    \includegraphics[width=\columnwidth]{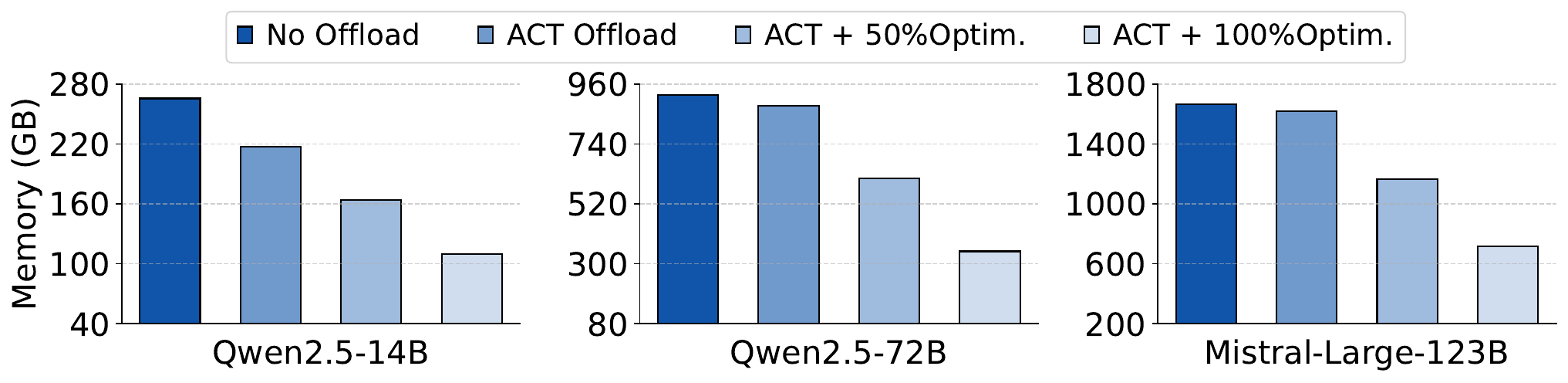}
    %\vspace{-0.2cm}
    \includegraphics[width=\columnwidth]{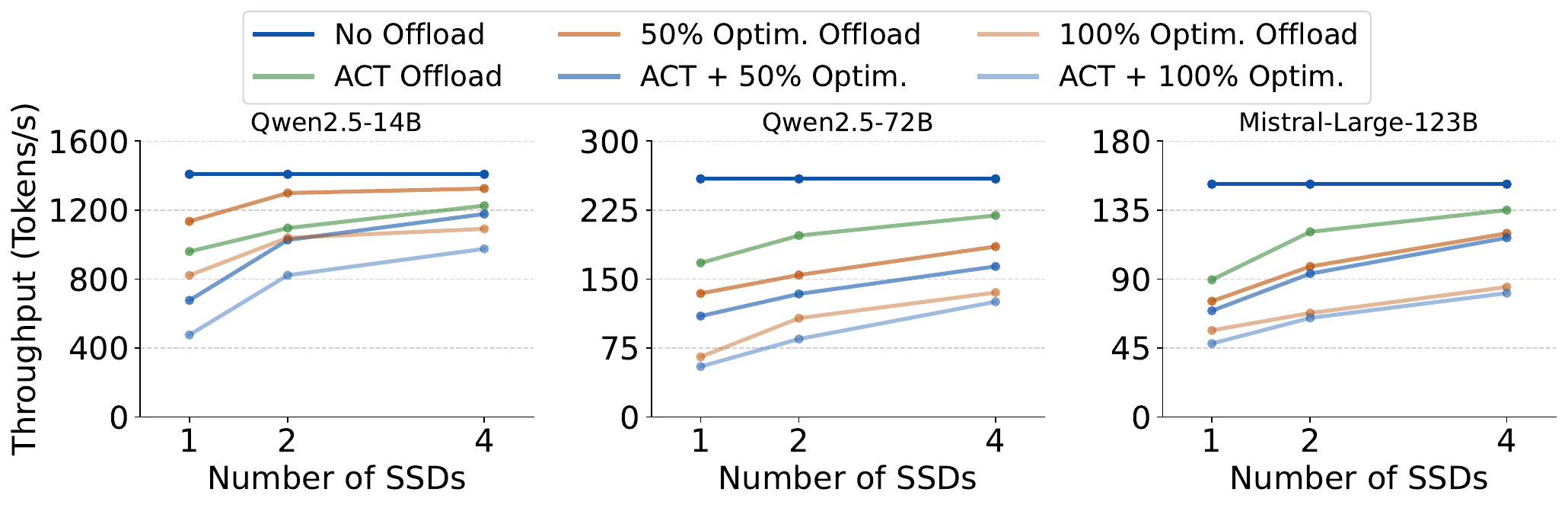}
    \vspace{-0.6cm}
    \caption{Performance comparison of NVMe SSD counts.}
    \label{fig:nvme_count_comparison}
\end{figure}

\subsection{Analysis of NVMe Offloading}
\label{sec:nvme_analysis}

For models exceeding CPU memory capacity, SlideFormer leverages the optional NVMe tier.  Activations and optimizer states can be offloaded asynchronously, with support for GPUDirect Storage and configurable offload fractions (50\% or 100\%) for optimizer states. Figure~\ref{fig:nvme_count_comparison} illustrates the trade-off between the CPU memory savings achieved through various offloading strategies and the corresponding impact on throughput: First, performance scales near-linearly with the number of NVMe drives, as I/O bandwidth becomes the primary bottleneck. Second, by enabling all offloading options, SlideFormer can reduce CPU memory consumption by 60-80\%, with a corresponding throughput degradation contained within 30-50\%. Third, the optimal offloading strategy is model-size dependent. For smaller models like Qwen2.5-14B, activations constitute a larger portion of the offloaded data. Offloading them provides significant memory savings but incurs a notable performance penalty as it impacts both the forward and backward passes. In this case, offloading optimizer states alone yields a better performance-to-memory trade-off. Conversely, for larger models where optimizer states dominate the memory footprint, offloading them first is most effective, and the additional, marginal impact of offloading activations becomes negligible. We therefore recommend offloading activations only for the largest models or under severe CPU memory constraints.

\begin{figure}[htbp]
    \centering
    \includegraphics[width=0.75\columnwidth]{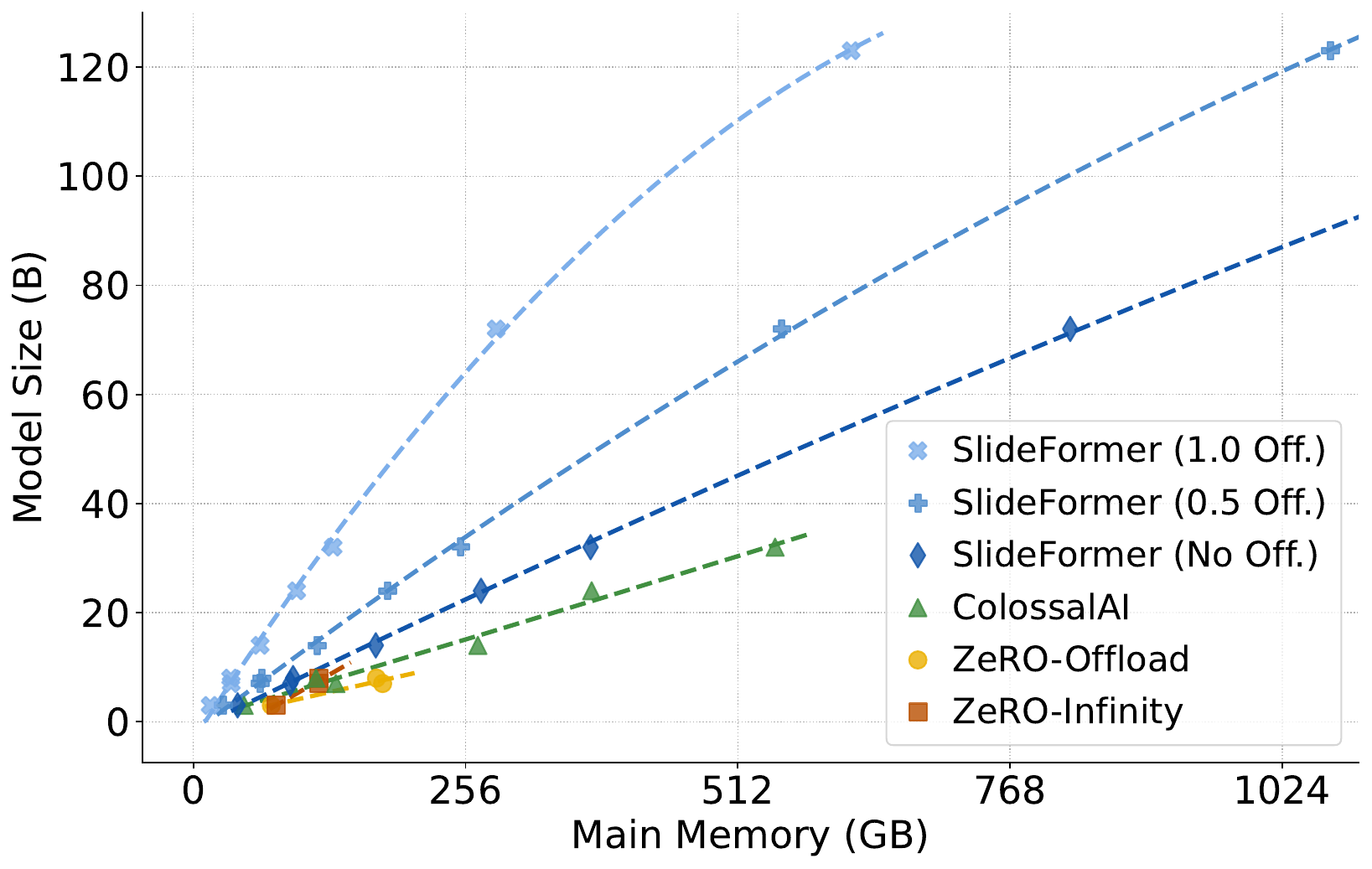}
    \vspace{-0.3cm}
    \caption{Maximum trainable model size.}
    \label{fig:max_model_size}
\end{figure}

\subsection{Maximum Trainable Model Size}
Figure~\ref{fig:max_model_size} presents a comparison of the maximum model sizes that can be fine-tuned using SlideFormer versus baseline frameworks, and each point is derived from actual tests conducted on listed pre-trained models. The experimental results demonstrate that, unlike other baselines which are constrained by GPU memory and thus limited in max trainable model size (e.g., Zero-offload supports up to 8B parameters, and ColossalAI supports up to 32B parameters), SlideFormer significantly extends the upper limit of fine-tunable model sizes. By shifting the primary memory constraint to CPU memory, SlideFormer enables the fine-tuning of models exceeding 123B parameters on a single GPU. For a high-end PC equipped with 256GB of CPU memory, enabling NVMe offloading allows fine-tuning models up to 90B parameters and can fine-tune models within 24B without throughput loss, as shown in Figure~\ref{fig:qwen_comparison}.

\subsection{Compared to Related Works}
\begin{figure}[!htbp]
    \centering % 居中整个 figure 环境的内容
    %\includegraphics[width=\linewidth]{figure/performance_comparison_gpt2_13b.pdf}
    %\vspace{-0.1cm}
    \includegraphics[width=\linewidth]{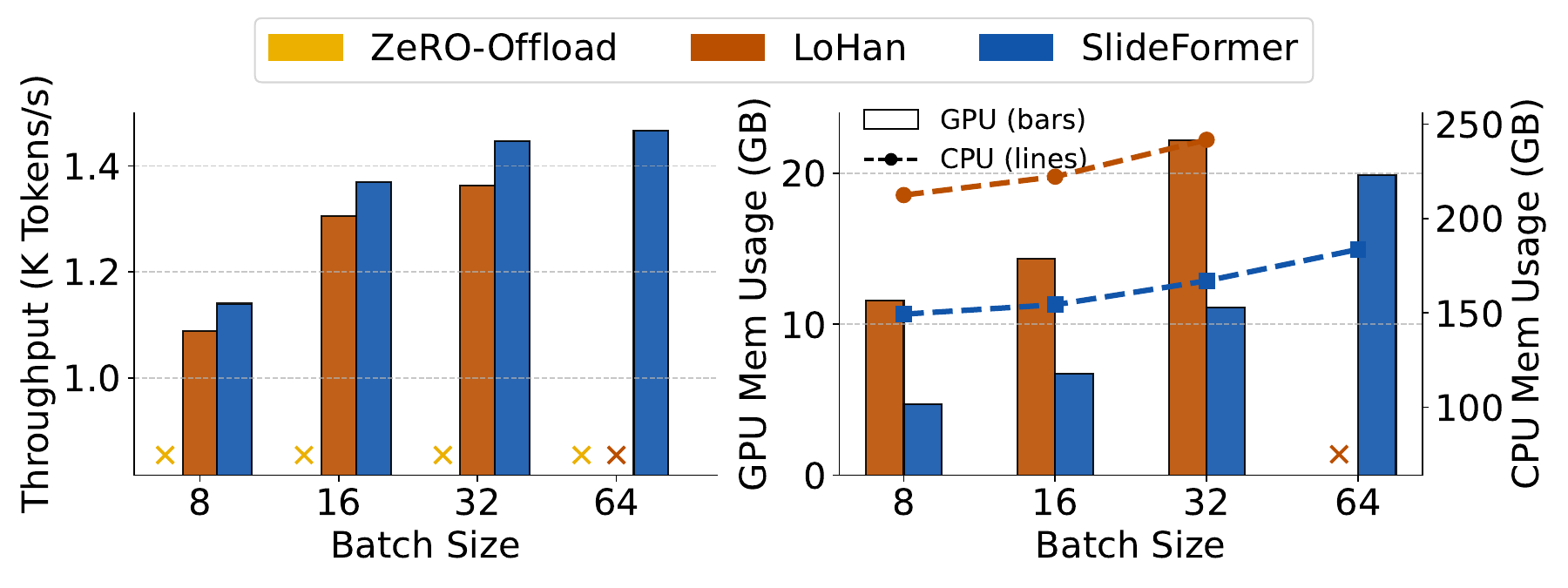}
        %\subcaption{Memory Usage Comparison} % 子图标题
        %\label{fig:memory} % 子图标签，用于引用
    \vspace{-0.7cm}
    \caption{Throughput and memory comparison between SlideFormer and LoHan for GPT2-13B on RTX4090.} % 整组图片的总标题
    \label{fig:vs_gpt2} % 总图标签
\end{figure}
In recent research, LoHan~\cite{liao2024lohanlowcosthighperformanceframework} is one of comparable to our work. However, it only supports GPT-2 and uses a non-standard loss function (MSE) during evaluation to sidestep the associated GPU memory overhead. Figure~\ref{fig:vs_gpt2} shows that under a standard GPT-2 Fine-tuning task, SlideFormer achieves superior performance, delivering higher throughput and consuming < 50\% of the GPU memory and saving ~30\% in CPU memory usage. ZeRO-Offload failed to run due to exceeding GPU memory. This result fundamentally validates our better architecture design and memory management compared to LoHan, which make SlideFormer the current optimal co-designed solution for current single GPU fine-tuning tasks.

\section{Conclusion}
In this paper, we present SlideFormer, a novel system that implements a holistic heterogeneous co-design, which significantly enhances the efficiency of full-parameter LLM fine-tuning on a single GPU. SlideFormer achieves 1.40-6.27× throughput gains while substantially halving CPU/GPU memory usage. It enables training 6× larger models and handling 8× larger batch sizes, demonstrating high compatibility (over 95\% peak performance on both NVIDIA\&AMD GPUs) with the latest LLMs. The primary significance of SlideFormer is its democratization of LLM fine-tuning, empowering individual researchers and smaller organizations.

%Looking ahead, several areas offer opportunities for further improvement. Current NVMe offloading incurs a 30-50\% performance overhead for the largest models and raises SSD lifespan concerns. Future efforts will focus on optimizing these offloading strategies and developing more intelligent, layer-aware scheduling algorithms to minimize impact. Support will also be extended beyond conventional Transformer architectures to include emerging structures like Mixture of Experts (MoE). Finally, integrating established memory-saving techniques such as quantization and Parameter-Efficient Fine-Tuning (PEFT) will be prioritized to further decrease resource demands.
% \begin{acks}
% \end{acks}

\bibliographystyle{ACM-Reference-Format}
\bibliography{my-base}

%\appendix

\end{document}